\documentclass[letter]{aa}  

\usepackage{graphicx}
\usepackage{txfonts}
\newcommand{\civ}{\hbox{C\,{\sc iv}\,$\lambda1550$}}
\newcommand{\ciii}{\hbox{C\,{\sc iii}]$\lambda1908$}}
\newcommand{\heii}{\hbox{He\,{\sc ii}\,$\lambda1640$}}
\newcommand{\nv}{\hbox{N\,{\sc v}\,$\lambda1240$}}
\newcommand{\siiv}{\hbox{Si\,{\sc iv}\,$\lambda1398$}}
\begin{document}

   \title{A candidate quadruple AGN system at $z\sim3$}

   \subtitle{}

   \author{Eileen Herwig
          \inst{1}
          \and 
          Fabrizio Arrigoni Battaia\inst{1}
          \and
          Eduardo Ba\~nados\inst{2}
          \and 
          Emanuele P. Farina\inst{3}
          }

   \institute{Max-Planck-Institut f\"ur Astrophysik, Karl-Schwarzschild-Straße 1, D-85748 Garching bei M\"unchen, Germany\\
              \email{eherwig@mpa-garching.mpg.de}
              \and
             Max Planck Institut f\"ur Astronomie, K\"onigstuhl 17, D-69117, Heidelberg, Germany
             \and 
             International Gemini Observatory, NSF’s NOIRLab, 670 N A’ohoku Place, Hilo, Hawai’i 96720, USA
              }

   \date{Received ; accepted }

 
  \abstract
  {Multiple galaxies hosting active galactic nuclei (AGNs) at kpc separation from each other are exceedingly rare, and in fact, only one quadruple AGN is known so far. These extreme density peaks are expected to pinpoint protocluster environments and therefore be surrounded by large galaxy overdensities. In this letter, we present another quadruple AGN candidate at $z \sim 3$ including two SDSS quasars at a separation of roughly 480~kpc. The brighter quasar is accompanied by two AGN candidates (a type 1 AGN and a likely type 2 quasar) at close ($\sim 20$~kpc) separation identified through emission line ratios, line widths and high ionization lines like \nv. The extended Ly$\alpha$ emission associated with the close triple system is more modest in extent and brightness compared to similar multiple AGN systems and could be caused by ram-pressure stripping of the type-2 quasar host during infall into the central dark matter halo. The predicted evolution of the system into a $z=0$ galaxy cluster with the AGN host galaxies forming the brightest cluster galaxy needs to be further tested by galaxy overdensity studies on large scales around the quadruple AGN candidate. If confirmed as a quadruple AGN with X-ray observations or rest-frame optical line ratios,  this system would represent the second AGN quartet, the highest-redshift multiplet and the closest high-redshift triplet known.}


   \maketitle
%

\section{Introduction}
In recent years, efforts to identify the progenitors of galaxy clusters, so called protoclusters (e.g., \citealt{Overzier2016}), have accelerated. 
Protoclusters in simulations are simply defined as structures eventually coalescing into a virialized galaxy group of mass $M > 10^{14}\ {\rm M}_{\odot}$ at $z=0$. Observationally, the identification of protoclusters is more complicated. 
High-$z$ quasars have long been believed to be signposts of peaks in the density distribution, but remain controversial as a method to identify protoclusters \citep{Uchiyama18, Habouzit2019}. While they often do reside in overdensities \citep{Fossati2021, Garcia-Vergara2022}, their typical halo masses of ${\rm 10^{12.5}\ M_{\odot}}$ at $z \sim 3$ are much lower than cluster masses \citep{White12, Trainor12, Husband13}, requiring substantial growth to form a galaxy cluster until $z = 0$.

\begin{figure*}
   \resizebox{\hsize}{!}
            {\includegraphics[width=0.1\textwidth]{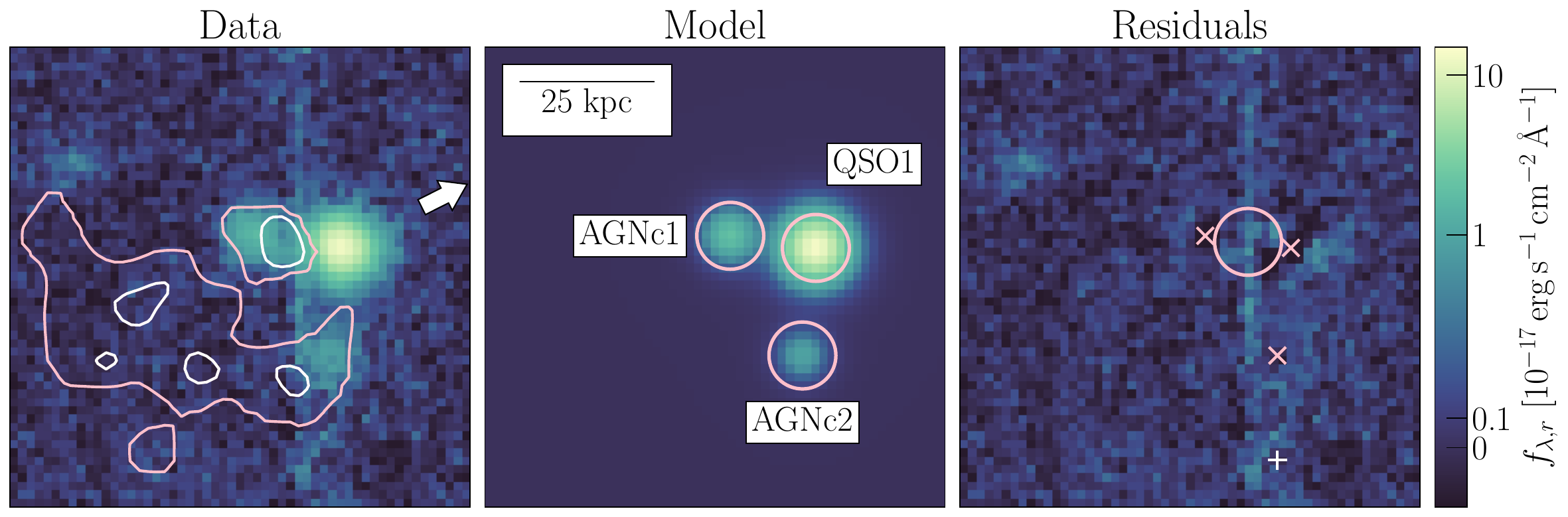}}
      \caption{\textbf{Left:} Extracted MUSE $r$-band map with overlayed the 2$\sigma$ and 5$\sigma$ isophotes (pink) of extended Ly$\alpha$ emission associated with the system \citep{Herwig2024}. Shown is a cut-out of size $11\arcsec \times 11\arcsec$.
      \textbf{Middle:} Best fitting Moffat model of the continuum sources. Pink circles indicate the apertures used for spectral extraction (Fig. \ref{fig:spectra}). \textbf{Right:} Residual flux density after subtracting the model. The crosses show the centroid position of subtracted sources. The pink circle indicates the aperture used to calculate the maximum apparent magnitude of an interloping galaxy (if any). The white cross shows the position of the spectrum used to evaluate the amount of CCD gap contamination (App.~\ref{sec:appcontamination}).}
         \label{fig:continuum}
\end{figure*}
While single bright quasars may pinpoint protocluster cores, physically associated multiple active galactic nuclei (AGN) systems could be even more reliable tracers of local overdensities \citep{Onoue2018, Sandrinelli2018}, albeit also not a firm indicator \citep{Fukugita2004, Green2011}, as a number of highly active, massive galaxies require ample supply of cold gas feeding supermassive black holes (SMBHs) and star formation. In the framework of hierarchical structure formation, such objects are most likely to reside in peaks of the density field populating the rare massive end of the halo mass function \citep{Bhowmick2020}. In the absence of dense environments, multiple AGN systems would be exceedingly rare due to their transient nature \citep{Hennawi2006}.
Interactions of three SMBHs are also predicted to be almost always necessary to form ultra-massive black holes with black hole masses $M_{\rm BH} > 10^{10}\ {\rm M}_{\odot}$ \citep{Hoffman2023}, further highlighting their importance in cluster formation as the central brightest cluster galaxies commonly host ultra-massive black holes (e.g., \citealt{Dullo2017, Mehrgan2019}).

So far, only a few high-z multiple AGN systems on kiloparsec scales are known. These objects are likely in an early stage of galaxy merger and might share a dark matter halo, but not their host galaxies. In contrast to that, sub-arcsecond triple AGNs trace local late-stage mergers with multiple cores in the same host galaxy (e.g., \citealt{Liu11, Kalfountzou17, Pfeifle19}). This work concentrates on the former AGN multiplets on kiloparsec scales, of which only five are known.
Specifically, \cite{Hennawi15} report the discovery of a quasar quartet at the peak of cosmic noon, $z \sim 2$. The AGN are embedded in a massive Ly$\alpha$ nebula extending by more than 300~kpc, indicative of a substantial reservoir of cool ($10^4$~K) gas (${\rm \sim 10^{11} M_{\odot}}$). The AGN are all found within a sphere of projected radius <150~kpc. The first triple quasar to be identified also resides at a redshift of 2 within a projected distance of less than 50~kpc, consistent with the typical onset of AGN activity in galaxy mergers \citep{Djor07}. Another triplet has been detected at $z \sim 1.5$, consisting of a close pair with a companion at 200~kpc projected distance \citep{Ema13}. The third triplet known is likewise embedded in a massive Ly$\alpha$ nebula at $z \sim 3$ and comprises two quasars and one faint AGN at projected distances below 100~kpc \citep{FAB2018}. A fourth triplet was found in a sample of almost 20,000 AGN candidates at a redshift of 1.13 \citep{Assef18}.

Here, we report on the discovery of a quadruple AGN candidate at $z \sim 3$ consisting of a close triplet, contained within 20~kpc in projection, and a forth quasar at a separation of 478~kpc. 
Throughout this work, we assume a flat ${\rm \Lambda}$CDM cosmology with ${\rm H_0\ =\ 67.7\ km\, s^{-1}\, Mpc^{-1}}$, ${\rm \Omega_m\ =\ 0.31}$ and ${\rm \Omega_{\Lambda}\ =\ 0.69}$ \citep{Planck18cosmo}. Reported quasar magnitudes are in the AB system.

\section{Observation and data analysis}

One quasar in the system, SDSS~J101254.73+033548.7 (hereinafter QSO1), has been targeted with the Multi Unit Spectroscopic Explorer (MUSE; \citealt{Bacon2010}) on the Very Large Telescope as part of a survey to study the environment of quasar pairs \citep{Herwig2024}. 
QSO1 is a known member of a quasar pair from the SDSS DR12 \citep{Paris17}. The projected distance to the second quasar in the pair SDSS~J101251.06+033616.5 (hereinafter QSO2) is 478~kpc. For QSO2, only the SDSS spectrum is available. In SDSS DR17 \citep{SDSSDR17}, the $r$-band magnitude of the quasars integrated within the SDSS fiber with a diameter of 3$\arcsec$ are $m_{\rm QSO1,SDSS} = (21.89 \pm 0.10)$~mag and $m_{\rm QSO2} = (21.65 \pm 0.09)$~mag.
Here we briefly describe the observations and data reduction.
More detailed descriptions of sample selection, observation and subsequent data analysis can be found in \cite{Herwig2024}. The MUSE data were acquired in Wide Field Mode on March 14, 2018 (seeing of $0.88\arcsec$), yielding a field of view (FoV) of $1\arcmin \times 1\arcmin$ (pixel scale of $0.2\arcsec$) and targeting only QSO1 as the separation of the pair is larger than the MUSE FoV. 
Three exposures were taken with ${\rm 880\, s}$ on-source time each, rotated by 90~degrees with respect to each other and covering a wavelength range of $4750\, \AA - 9350\, \AA$ with a sampling of $1.25\, \AA$. At the wavelength of Ly$\alpha$, the spectral resolution of $R \approx 1815$ corresponds to ${\rm FWHM} = 166$~km~s$^{-1}$ (${\rm FWHM} = 95$~km~s$^{-1}$ at \ciii).
The data have been reduced following \cite{Farina19} and \cite{GonzalezLobos23}. Specifically, we use the MUSE pipeline v 2.8.7 \citep{Weilbacher14, Weilbacher16, Weilbacher20} to perform bias- and sky-subtraction, flat fields, twilight and illumination correction and wavelength and flux calibration. To remove residual sky emission lines, we use the Zurich Atmospheric Purge \citep{SOTO16}. After masking gaps between the CCDs and correcting offsets between the exposures with custom routines, the three exposures are median combined to obtain the final data cube. To take into account correlated noise introduced during data processing, the variance cube produced by the pipeline is rescaled layer-by-layer to the RMS of a background region in the science cube, resulting in a mean pixel-wise noise level at the position of QSO1 of ${\rm 4.6 \times 10^{-20}\ erg\, s^{-1}\, cm^{-2}\, \AA^{-1}}$. 

\section{Results}
\subsection{Discovery of a candidate quadruple AGN}
\label{subsec:discovery}

\begin{figure}
   \centering
   \includegraphics[width=\hsize]{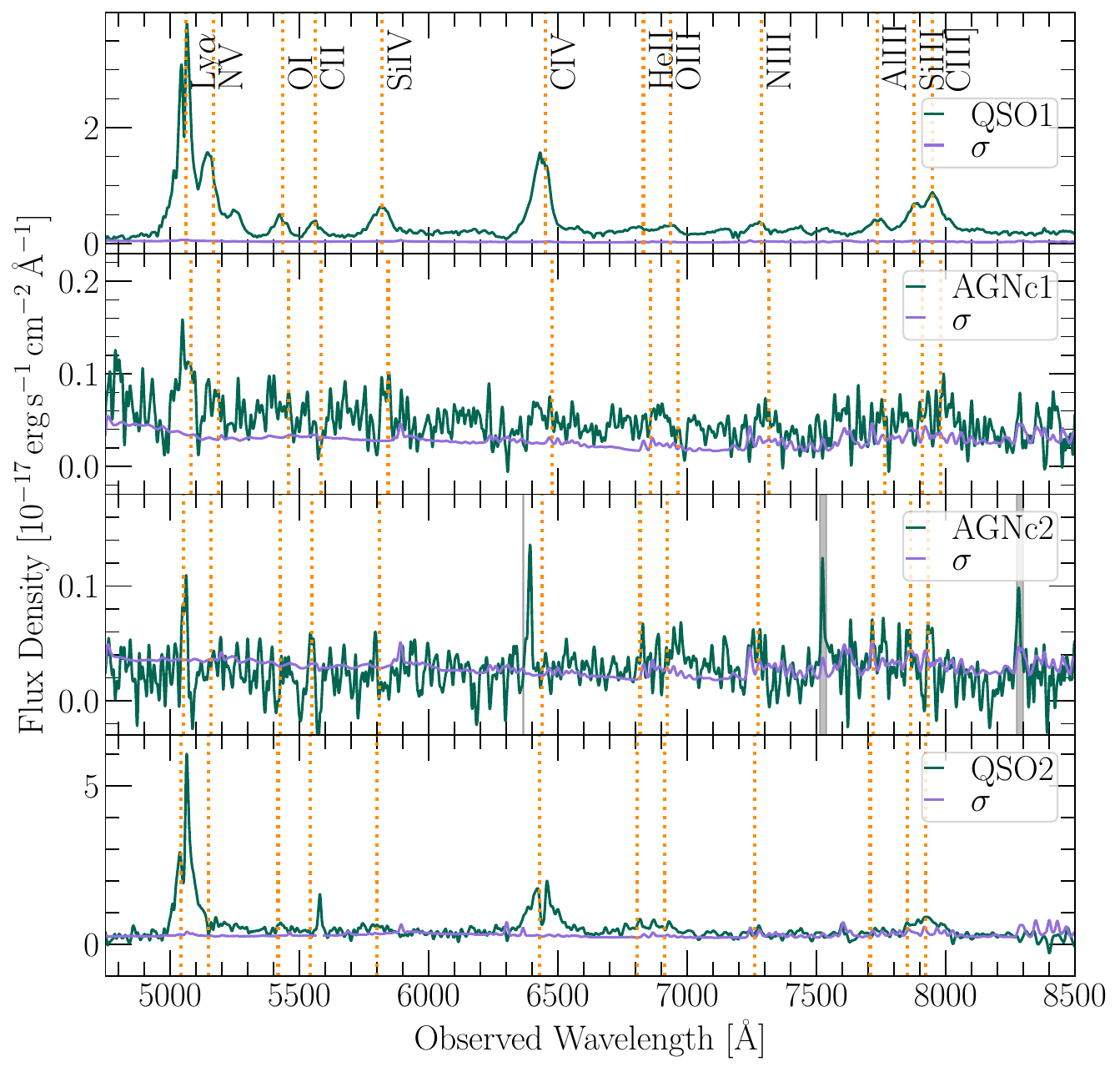}
      \caption{One dimensional spectra for the candidate quadruple AGN system. QSO1 and QSO2 are the two SDSS quasars, while AGNc1 and AGNc2 are the two additional AGN candidates. For each source, we indicate both the data and noise spectrum (see legend). Orange dashed vertical lines indicates the location of important rest-frame UV transitions at the redshift of each object obtained from a fit to the \ciii\ line. Location of sky-lines with residuals in the spectrum of AGNc2 are indicated with shaded vertical areas. Due to the vastly different magnitude of the objects, the y-axis range differs between panels.}
         \label{fig:spectra}
   \end{figure}

By collapsing the final MUSE cube along the wavelength axis, we obtained a continuum image of the field surrounding QSO1. 
In this image (r-band in Fig.~\ref{fig:continuum}, left), two additional continuum sources are visible in close proximity to QSO1. While similar geometries can be seen as a result of gravitational lensing of a single source, we can confidently reject this possibility in this case (Appendix~\ref{sec:gravlens}). We instead identify both of these sources (hereinafter called AGNc1 and AGNc2) as AGN candidates as detailed in the following. To constrain their nature, we extracted spectra from the MUSE cube within an aperture of 0.8$\arcsec$ to minimize contamination from each other's PSFs. To determine their positions, we first modeled the PSF of a star within the MUSE FoV with a 2D Moffat function and fit the model to the three continuum sources observed with MUSE (QSO1, AGNc1, AGNc2) by fixing the FWHM and leaving amplitude and position free to vary (Fig.~\ref{fig:continuum}, middle panel). The peak of the Moffat function was used as center of the aperture for spectral extraction and is indicated with pink circles (crosses) in the middle (right) panel of Fig.~\ref{fig:continuum}. 
To reduce noise while retaining the signal, the spectra were smoothed by a 1D Gaussian kernel with $\sigma=3$~pixels, corresponding to $3.75\ \AA$. Before smoothing the spectrum of AGNc2, we masked a sky emission line at 6365.3~$\AA$ as it would otherwise be convolved with the \civ\ line (Appendix~\ref{sec:appcontamination}). The spectra are shown in Fig.~\ref{fig:spectra}, together with the SDSS spectrum of QSO2. The two QSOs and two AGN candidates are all detected in Ly$\alpha$, confirming that they reside at roughly the same redshift. AGNc1 tentatively shows a broad \ciii\ line and has excess flux in \nv, \siiv\ and \heii, while in the spectrum of AGNc2, there is a tentative detection of a narrow \ciii\ and \heii\ line as well as a prominent narrow, blue-shifted \civ\ line. 
Within the aperture of $0.8\arcsec$ used for spectral extraction, the $r$-band magnitude for the two AGN candidates is $m_{\rm AGNc1} = (24.36 \pm 0.49)$~mag and $m_{\rm AGNc2} = (24.74 \pm 0.74)$~mag\footnote{In this smaller aperture, the magnitude of QSO1 decreases to $m_{\rm QSO1} = (22.34 \pm 0.09)$~mag}. 
Between QSO1 and AGNc1, we measured an angular separation of 2.1$\arcsec$, corresponding to the distance $\Delta d = 16$~kpc at this redshift. QSO1 and AGNc2 are separated by 2.6$\arcsec$ or 20~kpc, and QSO2 is 61.6$\arcsec$ or 478~kpc away from QSO1.
We obtained the systemic redshift $z_{\rm CIII]}$ by fitting the \ciii\ line with a double Gaussian model and using the peak of the line complex as reference wavelength. 
The value obtained for AGNc1 is likely overestimated as a redshift of $z = 3.181$ implies a highly unusual Ly$\alpha$ line shape dominated by the blue wing (Fig.~\ref{fig:fitagn2}). If this was the case, the \heii\ line might be misidentified and possibly not detected. Positions of the AGN candidates and projected distance between each other as well as redshift and velocity shifts are summarized in Table \ref{table:agn}.

We fit the spectra as described in Appendix~\ref{sec:appfit} and summarize the resulting line fluxes, widths and upper flux limits in Table~\ref{table:flux}.
The line width of AGNc2 is narrow with ${\rm FWHM_{CIII]}} = 600^{+80}_{-70}$~km~s$^{-1}$ and consistent with a Type II AGN, while the spectrum of AGNc1 is dominated by broad lines of ${\rm FWHM_{CIII]}} = 5040^{+1390}_{-3790}$~km~s$^{-1}$ indicative of a Type I AGN.

\subsection{Properties of QSO1 and QSO2}

Using the fitting results (Appendix~\ref{sec:appfit}), we can derive additional properties for the brighter SDSS quasars QSO1 and QSO2. We calculated the black hole masses using the relation and absolute accuracy estimate derived for \civ\ in \cite{Vestergaard2006} and obtained $\log M_{\rm BH, QSO1} = (8.3 \pm 0.56)\ {\rm M_{\odot}}$ and $\log M_{\rm BH, QSO2} = (8.8 \pm 0.56)\ {\rm M_{\odot}}$.
At the position of QSO1, radio emission is detected with a flux density of $f_{3 {\rm GHz}} = (0.7\, \pm\, 0.1)\ {\rm mJy\ beam^{-1}}$ in VLASS \citep{Lacy2020} and $f_{1.4 {\rm GHz}} = (1.02\, \pm\, 0.15)\ {\rm mJy\ beam^{-1}}$ in VLA FIRST \citep{Becker94}, leading to a radio slope of $\alpha = -0.49$ ($f_{\nu} \propto \nu^{\alpha}$). 
Additionally using the UV continuum slope from the fit to the spectrum of QSO1 (Appendix~\ref{sec:appfit}) of $\beta = -1.04$ ($f_{\lambda} \propto \lambda^{\beta}$), we calculated the ratio $R = f_{5 {\rm GHz}}/f_{4400 \AA}$ \citep{Kellermann1989} by extrapolating the rest-frame flux densities and obtained $R=3078$, classifying QSO1 as radio-loud. Although no additional radio emission is detected, we cannot exclude that AGNc1 and AGNc2 are radio-loud due to the shallow surveys depth\footnote{The VLASS continuum image rms is $70 \mu$Jy/beam \citep{Lacy2020}, FIRST data has an rms of $149 \mu$Jy/beam close to the system location\citep{Becker94}.} and their dim UV continuum.

\begin{table*}
\caption{Quasar and AGN candidate properties and distances between each other.}             
\label{table:agn}      
\centering          
\begin{tabular}{c c c c c c c }
\hline\hline       
Object& RA (J2000)\tablefootmark{a} & Dec (J2000)\tablefootmark{a} & $r$ [mag]\tablefootmark{b} & ${z_{\rm CIII]}}$ & $\Delta v$ to QSO1 [${\rm km\, s^{-1}}$] & $\Delta d$ to QSO1 [\arcsec/kpc]\\ 
\hline                    
   QSO1 & 10:12:54.74 & +03:35:48.93 & $21.89 \pm 0.10$\tablefootmark{c} & 3.1638 $\pm$ 0.0004 & 0 &0/ 0\\  
   AGNc1 & 10:12:54.88 & +03:35:49.22 & $24.36 \pm 0.49$ & 3.181 $\pm$ 0.008 & -1590 $\pm$ 766 & 2.1/16\\
   AGNc2 & 10:12:54.77 & +03:35:46.36 & $24.74 \pm 0.74$ & 3.156 $\pm$ 0.001 & 761 $\pm$ 166 & 2.6/20\\
   QSO2 & 10:12:51.07\tablefootmark{d} & +03:36:16.56\tablefootmark{d} & $21.65 \pm 0.09$\tablefootmark{c} & 3.150 $\pm$ 0.001 & 1347 $\pm$ 188 & 61.6/478\\
\hline                  
\end{tabular}
\tablefoot{
\tablefoottext{a}{Coordinates determined from best fitting Moffat model}
\tablefoottext{b}{Calculated in the MUSE data within an aperture diameter of 1.6\arcsec}
\tablefoottext{c}{SDSS magnitude within an aperture of 3\arcsec}
\tablefoottext{d}{Coordinate taken from SDSS}
}
\end{table*}

\subsection{Emission line diagnostics}

\begin{figure}
   \centering
   \includegraphics[width=\hsize]{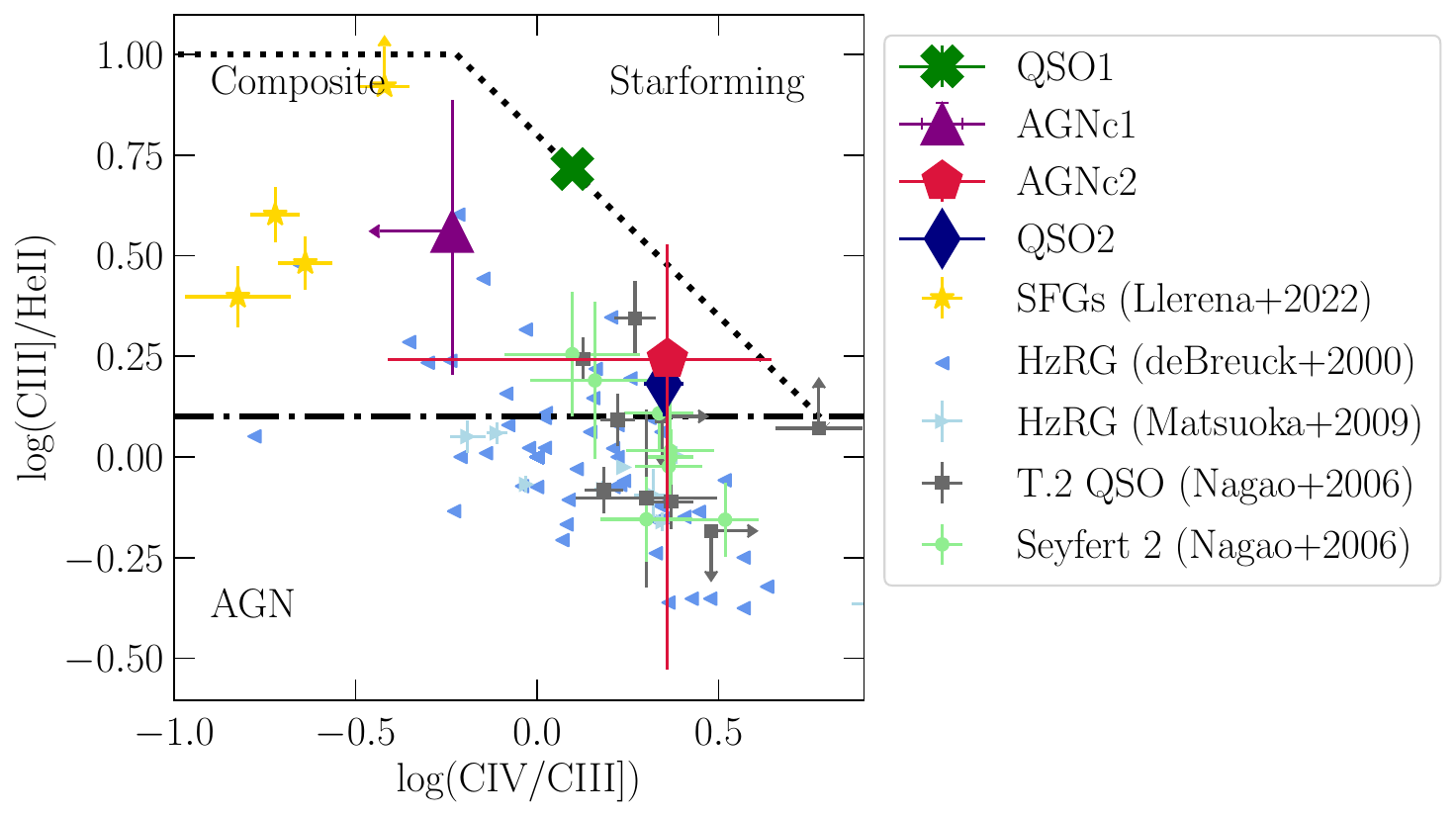}
      \caption{Diagnostic diagram based on the line ratios \civ/\ciii\ and \ciii/\heii. Demarcation lines between AGN, star forming galaxies and composite line ratios are taken from \cite{Hirschmann2019}. The values for QSO1, AGNc1, AGNc2 and QSO2 are compared to literature values for different types of galaxies: \ciii-selected star forming galaxies at $z > 2.9$ binned by stellar mass \citep{Llerena2022}, high-$z$ radio galaxies with detections in all lines \citep{deBreuck2000, Matsuoka2009}, high-$z$ X-ray selected type 2 quasars \citep{Nagao2006} and local Seyfert 2 AGNs \citep{Nagao2006}.
      }
         \label{fig:diadia}
   \end{figure}

A commonly adopted diagnostic diagram to distinguish AGNs from star forming galaxies (SFGs) in rest-frame UV uses the line ratios \civ/\ciii\ (C43), mainly influenced by the ionization parameter, and \ciii/\heii\ (C3He2), which is sensitive to metallicity and allows distinction between shocks and AGN photoionization \citep{Feltre2016}. \cite{Hirschmann2019} propose demarcation lines to distinguish between active and inactive galaxies, but current models are not capable of fully reproducing observed line ratios, leading to large impurities in the selection of AGNs and specifically faint AGNs. Classification is further complicated because certain types of galaxies would not appear in such a diagnostic diagram: high-$z$ radio galaxies (HzRGs) commonly have no detection of \civ\ (56~\% in \citealt{deBreuck2000}) and SFGs usually show \civ\ and \siiv\ in absorption instead of emission (e.g. \citealt{Shapley2003, Calabro2022}).
In Fig.~\ref{fig:diadia} we show the line ratios of the candidate AGNs and SDSS quasars together with line ratios of SFGs and different types of AGNs from the literature. Both quasars and both AGN candidates as well as 30~\% of the literature AGNs fall within the composite region of the diagram, indicating contribution from both star formation and AGN activity. Surprisingly, the line ratios of QSO1 indicate the highest contribution from star formation of all four objects although it is definitively classified as quasar, further highlighting the difficulty in AGN selection based on UV line ratios. It is also notable that the stacks of observed SFGs selected based on \ciii\ emission are classified as composite objects as well.

However, the detection of \nv\ and \siiv, two highly ionized lines, and the extreme line width of AGNc1 larger than $5000\ {\rm km\, s^{-1}}$ firmly point toward an AGN origin. If these broad lines were produced by galactic-scale outflows instead, the violent kinematics should lead to a high contribution from shocks and likely lead to a detectable \civ\ emission line. The upper limit on the EW of \civ\ lies above the AGN demarcation line of $\lesssim 3 \AA$ at the corresponding \civ/\heii\ ratio limit for AGNc1 of <~1 \citep{Nakajima2018} and AGN ionization can therefore not be excluded based on the non-detection of the line. The ratio of \siiv\ to \heii\ exceeds those typical for HzRGs \citep{Humphrey2008} as well as quasars \citep{Nagao2006}. On the other hand, the \civ\ EW of AGNc2 lies firmly above the demarcation line of 12~\AA\ in \cite{Nakajima2018} at the corresponding \civ/\heii\ ratio for AGNc2, indicating that this source is an AGN.

\section{Discussion}

\subsection{Clues from the peculiar extended Ly$\alpha$ emission}

The pink contours in Fig. \ref{fig:continuum} show the 2$\sigma$ and 5$\sigma$ levels of extended Ly$\alpha$ emission, corresponding to a surface brightness of ${\rm 3.0 \times 10^{-18}\ erg\, s^{-1}\, cm^{-2}\, arcsec^{-2}}$ and ${\rm 7.4 \times 10^{-18}\ erg\, s^{-1}\, cm^{-2}\, arcsec^{-2}}$ respectively, and determined after subtracting empirical quasar point spread functions and continuum sources from the combined cube. This analysis is described in detail in \cite{Herwig2024} and \cite{GonzalezLobos23}.
The nebular emission extends for about 50~kpc in the south-east direction with respect to QSO1, starting from the projected location of AGNc2. Additional Ly$\alpha$ emission coincides with the position of AGNc1. Surprisingly, the surface brightness level of the extended Ly$\alpha$ emission is much lower than typically found for multiple AGN systems \citep{Hennawi15,FAB2018} and the AGN candidates are not embedded in the nebula. Instead, the flux-weighted centroid of the extended emission is offset from QSO1 by 34~kpc \citep{Herwig2024}.
Possible explanations for this could be a faster evolution and therefore early cool gas depletion of outlier density peaks in the universe or a later merger stage of the close AGN triplet candidate presented in this work. Moreover, strong feedback effects might have already depleted the circumgalactic medium of this system of cool gas.

The peculiar morphology of the extended Ly$\alpha$ emission could be further evidence for violent on-going galaxy interactions: if the ISM of AGNc2 is ram-pressure stripped during the infall into the hot halo of QSO1, the trailing dense gas behind AGNc2 could become visible as extended Ly$\alpha$ emission. 
Indeed, estimating the stripped gas mass yields a lower limit of at least $6.3 \times 10^9 {\rm M_{\odot}}$ when assuming the line-of-sight velocity, or roughly $8.6 \times 10^9 {\rm M_{\odot}}$ when using a conservative estimate of the 3D velocity, while the estimated cool gas mass visible as extended Ly$\alpha$ emission amounts to roughly $4.8 \times 10^9 {\rm M_{\odot}}$ (Appendix \ref{sec:app1}).
This first estimate confirms that ram-pressure stripping could be an effective mechanism in this configuration and could indeed explain the peculiar appearance of the extended Ly$\alpha$ emission.
Evidence for similar ram-pressure stripping events around high-z quasars have been recently invoked by other studies on extended Lya emission (e.g., \citealt{Chen2021}).

\subsection{Evolutionary scenario}

The close separation of QSO1, AGNc1 and AGNc2 and evidence for gas stripping and therefore ongoing galaxy interactions makes it very likely that the host galaxies will quickly merge into one massive galaxy. The result of such a merger is predicted to have a host halo mass of ${\rm 10^{13} M_{\odot}}$ \citep{Bhowmick2020}.
Three-body interactions have a high likelihood of ejecting one of the participants via gravitational recoil \citep{Partmann2024}. This is the most probable evolution of the close black hole system, although two or even three mergers are also possible and would lead to the emission of gravitational waves possibly contributing to the gravitational wave background \citep{Sesana2008} and likely the formation of an ultra-massive ($M_{\rm BH} > 10^{10}\ {\rm M}_{\odot}$) black hole. 
However, ejection or stalling of one or two black holes is predicted to not hinder fast growth: in simulations, the strong gravitational forces lead to quick infall times of gas into the most massive black hole and facilitate black hole mass growth up to $10^{10}\ {\rm M_{\odot}}$ irrespective of coalescence \citep{Hoffman2023}.
The extreme density peak implied by the presence of four AGNs in close proximity and three AGNs in the same dark matter halo is likely accompanied by a galaxy overdensity. This hypothesis needs to be tested by multi-wavelength observations targeting the galaxy population on Mpc scales around the quadruple AGN candidate. If confirmed, the most probable evolutionary path of this extraordinary system is the formation of a galaxy cluster in which the brightest cluster galaxy is formed out of the close triple AGN host galaxies.

\subsection{Incidence rate of AGN multiplets}

Following \citet{Hennawi15}, an estimate of the probability to find such two close companion AGNs to a known quasar can be gained from the two-point correlation function for quasars and, for a magnitude limit of 25~mag, amounts to $\sim 10^{-8}$, or $\sim 10^{-5}$ if considering any halo-scale triple AGN (Appendix~\ref{sec:app2pcf}). A different generalized estimate for the AGN multiplet incidence rate can be gained by considering the full observed parent sample.
Including the candidate presented in this work, 2 AGN multiplets have been identified within the QSO MUSEUM survey encompassing 134 quasars at $3<z<4$ observed with MUSE to similar depths (\citealt{FAB2019, Herwig2024}; Gonz\'alez Lobos et al. in preparation). This corresponds to an incidence rate of such systems of 1.5$^{+1.9}_{-1}$~\% with Poisson errors for small-numbers statistics from \cite{Gehrels1986ApJ}, much larger than the above calculations and than previously believed \citep{Hennawi15}. This implies $511^{+647}_{-341}$ undiscovered multiplet systems around SDSS quasars \citep{Paris2018} in the same redshift range (with a median redshift of $z = 3.226$), and, extrapolated to the entire sky, $2810^{+3559}_{-1873}$ AGN multiplet.
Assuming that black hole multiplets populate the massive end of the halo mass function and that 65 \% of them are visible as AGN multiplets \citep{Hoffman2023}, this number would amount to all halos down to $M_{\rm vir} = (5 \pm 1) \times 10^{13}$ M$_{\odot}$ of the $z = 3.226$ halo mass function \citep{Behroozi2013}.

\section{Summary}

We present an unlensed candidate quadruple AGN system at $z \sim 3$ consisting of two SDSS type 1 quasars at a separation of roughly 480~kpc (QSO1 and QSO2) and two AGN candidates (AGNc1 and AGNc2) accompanying QSO1 in close proximity with a projected separation of up to 20~kpc. The identification as probable AGN candidates is based on C43 versus C4He2 line ratios, high ionization lines like \nv\ in their spectrum and the emission line widths. 
AGNc1 is a potential type 1 AGN, while the spectral characteristics of AGNc2 align with those of a type 2 quasar. Further observations in X-ray or of rest-frame optical lines can confirm these objects as AGNs and allow for more reliable redshift determination.

The close triple system (QSO1, AGNc1, AGNc2) is associated with a Ly$\alpha$ nebula, although it is not embedded within it. A possible explanation for the presence of peculiar Ly$\alpha$-bright gas is ram-pressure stripping of AGNc2 during its infall into the halo of QSO1.

This extraordinary system likely pinpoints a site of galaxy cluster formation. This hypothesis needs to be tested further by studying the Mpc-scale galaxy overdensity surrounding the quadruple AGN candidate.

This new candidate quadruple AGN allow us to refine the incidence rate of such rare systems around $z\sim3$ SDSS quasars to 1.5$^{+1.9}_{-1}$~\%. Pre-selection of SDSS quasars with close faint sources (e.g., Euclid) is essential to build a statistical sample (12 systems), needed to confirm our statistics and constrain current models (e.g., \citealt{Hoffman2023}). Without it, additional $700^{+1566}_{-218}$ MUSE observations (1 hour/source) of SDSS quasars would be required. 

\begin{acknowledgements}
We thank the anonymous referee for their constructive comments that helped improve the manuscript and Guinevere Kauffmann for providing useful comments to an earlier version of this work. E.P.F. is supported by the international Gemini Observatory, a program of NSF NOIRLab, which is managed by the Association of Universities for Research in Astronomy (AURA) under a cooperative agreement with the U.S. National Science Foundation, on behalf of the Gemini partnership of Argentina, Brazil, Canada, Chile, the Republic of Korea, and the United States of America. 
\end{acknowledgements}

\bibliographystyle{aa}
\bibliography{lit.bib}
\begin{appendix}

\section{Rejecting the possibility of gravitational lensing}
\label{sec:gravlens}

A possible explanation for multiple quasars in close separation is gravitational lensing of a single source (e.g., \citealt{Agnello2018}).
In this case, the spectra would be very similar with slight variations due to different light paths. While all AGN in this system appear to have significantly different spectra (Fig.~\ref{fig:spectra}), this view might be skewed by the low signal to noise ratio of the fainter objects. While AGNc2 is a type 2 AGN candidate as it only shows narrow emission lines and should therefore be viewed at an entirely different angle than QSO1 and AGNc1, we explore the possibility that the latter two objects are actually two lensed images of the same quasar.

In this case, we expect to find a massive object capable of acting as the gravitational lens in between the two images. 
To calculate the necessary magnitude of a lensing elliptical galaxy, we followed \citet{Ema13}, making use of the Faber–Jackson relation by \citet{NigocheNetro10} and assuming an isothermal sphere to link mass and velocity dispersion. Assuming an Einstein ring radius of $\Theta = 1.05 \arcsec$, corresponding to half of the angular distance between QSO1 and AGNc1, the minimal possible apparent $r$-band magnitude of an interloper galaxy is $m = 21.6$ at $z \approx 1.89$, brighter than both the AGN candidates and QSO1.

We then inspected the residuals of the point spread function modeling (Fig. \ref{fig:continuum}, right). There, an artifact of the CCD gap becomes apparent as vertical line, but no foreground galaxy is revealed. The $r$-band magnitude in a 1.6$\arcsec$ aperture between QSO1 and AGNc1 (indicated as pink circle in Fig.~\ref{fig:continuum}, right) after masking negative values potentially arising due to the model subtraction is 26.4~mag, far below the faintest possible lensing galaxy.
Therefore we conclude that no lensing geometry fits the observed pattern without revealing the lensing galaxy in the data cube.

\section{Spectral fitting}
\label{sec:appfit}
We attempted to fit a set of rest-frame UV lines to the spectra of all four sources presented in Fig.~\ref{fig:spectra}, namely Ly$\alpha$, \nv, \siiv, \civ, \heii\ and \ciii. Due to the stark difference in magnitude and therefore data quality, we adopted two different methods.
All line fluxes, equivalent widths (EWs) and line widths are summarized in Table~\ref{table:flux}.

\begin{table*}
\caption{Line fluxes, equivalent widths and line width.}             
\label{table:flux}      
\centering          
\renewcommand{\arraystretch}{1.3}
\begin{tabular}{c c c c c c c c }     
\hline\hline       
Object& $f_{\rm Ly\alpha}$\tablefootmark{a} & $f_{\rm NV}$\tablefootmark{a} & $f_{\rm SiIV}$\tablefootmark{a}/ EW [\AA] & $f_{\rm CIV}$\tablefootmark{a}/ EW [\AA] & $f_{\rm HeII}$\tablefootmark{a} & $f_{\rm CIII]}$\tablefootmark{a} & ${\rm FWHM_{CIII]}}$\tablefootmark{b}\\ 
\hline                    
   QSO1\tablefootmark{c} & 352.8 $\pm$ 9.3 & 44.4 $\pm$ 1.5 & 40.2 $\pm$ 0.6 / $55 \pm 1$ & 112.3 $\pm$ 0.9 / $153 \pm 1$ & 17.3 $\pm$ 0.8 & 89.8 $\pm$ 0.9 & $5100 \pm 60$\\
   AGNc1\tablefootmark{d} & \multicolumn{2}{c}{6.0$^{+1.8}_{-2.1}$} & 1.4$^{+1.0}_{-0.3}$/ $30^{+43}_{-10}$ & <1.1/ <28 & 1.1$^{+0.3}_{-0.5}$ & 4.0$^{+2.4}_{-3.1}$ & $5040^{+1390}_{-3790}$\\
   AGNc2\tablefootmark{d} & 1.6$^{+0.7}_{-0.3}$  & <0.3 & <0.2/ <11 & 1.6$\pm 0.3$/ $61^{+28}_{-25}$ & 0.4$^{+0.7}_{-0.2}$ & 0.7$^{+0.3}_{-0.2}$ & $600^{+80}_{-70}$\\
   QSO2\tablefootmark{c} & 272.2 $\pm$ 9.1 & - & - & 159.9 $\pm$ 7.8/ $120 \pm 5$ & 47.1 $\pm$ 7.1 & $71.6 \pm 8.1$ & $5050 \pm 690$\\
\hline                  
\end{tabular}
\tablefoot{
\tablefoottext{a}{Line flux in ${\rm 10^{-17}\ erg\,s^{-1}\,cm^{-2}}$}
\tablefoottext{b}{Line width in ${\rm km\ s^{-1}}$, determined from \ciii\ and corrected for the Gaussian kernel smoothing applied before the fit}
\tablefoottext{c}{Values obtained from \texttt{pyQSOFIT}}
\tablefoottext{d}{Values obtained by Gaussian fitting}
}
\end{table*}

\subsection{SDSS QSOs}

The two SDSS detected quasars QSO1 and QSO2 have sufficient S/N to be fitted using the package \texttt{pyQSOFIT} \citep{Guo2018, Shen2019}. We used the spectrum extracted from the MUSE cube as explained in Sec.~\ref{subsec:discovery} for QSO1 as it has higher S/N than the SDSS spectrum. For QSO2, only the SDSS spectrum is available. 
Errors were estimated through 250 MCMC samplings, and the line width ${\rm FWHM_{CIII]}}$ was determined from the fit to the \ciii\ line. The Gaussian model used to describe the \ciii\ line is affected by the Gaussian smoothing kernel applied before fitting, and we thus corrected the values down to reflect the intrinsic line width.
Although we attempted to include a component for the \nv\ and \siiv\ line when fitting QSO2, we did not find a combination of priors that leads to convergence and we thus do not report fluxes for these lines.
From \texttt{pyQSOFIT} we also obtained the continuum luminosity $L(1350 \AA)$ and the FWHM of \civ\ and used them to calculate black hole mass estimates for QSO1 and QSO2 using the relation and absolute accuracy estimate derived in \cite{Vestergaard2006}. With $L(1350 \AA)_{\rm QSO1} = 10^{44.91 \pm 0.01}$~erg~s$^{-1}$, $L(1350 \AA)_{\rm QSO2} = 10^{45.32 \pm 0.02}$~erg~s$^{-1}$, ${\rm FWHM_{CIV,QSO1} = (3610 \pm 20)}$~km~s$^{-1}$ and ${\rm FWHM_{CIV,QSO2} = (5350 \pm 260)}$~km~s$^{-1}$, we obtained the values $\log M_{\rm BH, QSO1} = (8.3 \pm 0.56)\ {\rm M_{\odot}}$ and $\log M_{\rm BH, QSO2} = (8.8 \pm 0.56)\ {\rm M_{\odot}}$.

\subsection{AGN candidates}

Due to low S/N, the spectra of candidates AGNc1 and AGNc2 cannot be fitted with \texttt{pyQSOFIT} and we instead subtracted a constant local continuum value before fitting emission lines using the function \texttt{curve\_fit} in \texttt{scipy} \citep{2020SciPy}.

In particular, we obtained a first estimate of the line widths $\sigma_{\rm ini}$ by calculating the second moment of the Ly$\alpha$ line (AGNc1) and \civ\ line (AGNc2) in an unsmoothed spectrum. We used the resulting values of $\sigma_{\rm ini} = 1657^{+46}_{-57}$~km~s$^{-1}$ and $\sigma_{\rm ini} = 243^{+14}_{-18}$~km~s$^{-1}$ respectively (corresponding to ${\rm FWHM_{ini} = 3902^{+108}_{-134}}$ and ${\rm FWHM_{ini} = 572^{+33}_{-42}}$ respectively) to constrain the following fit.

\begin{figure}[h]
   \centering
   \includegraphics[width=0.49\hsize]{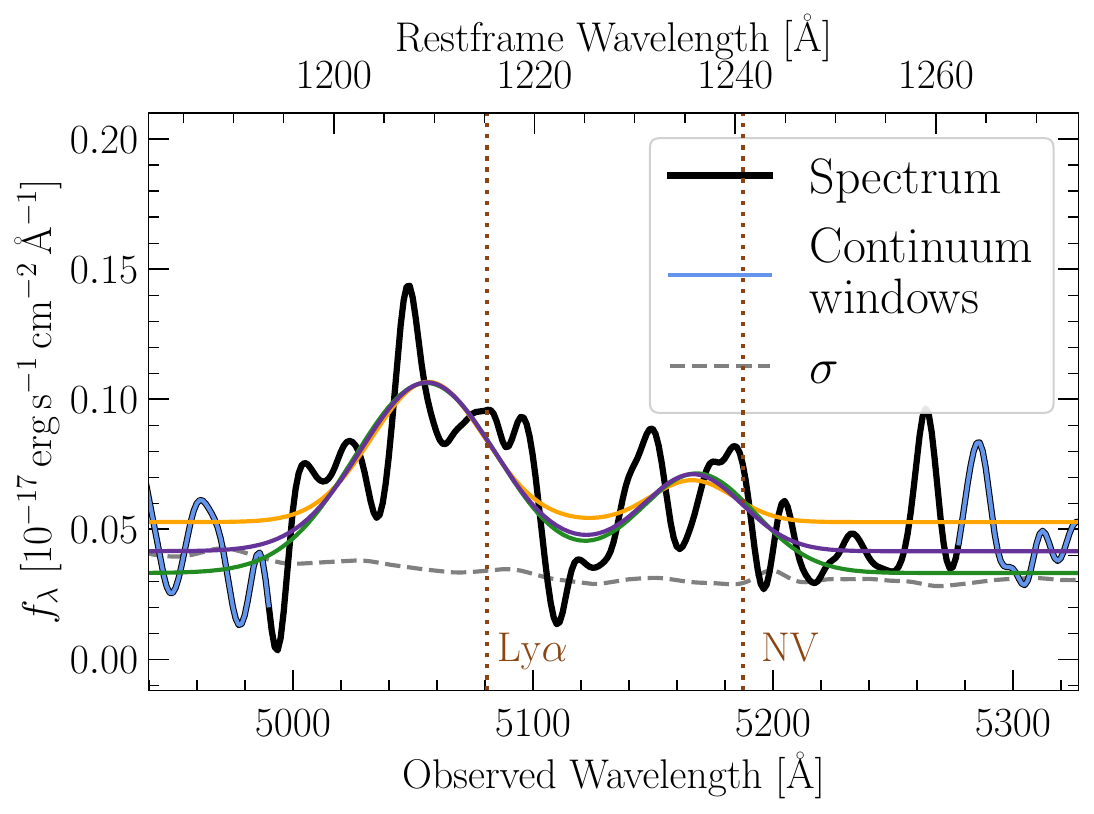}
   \includegraphics[width=0.49\hsize]{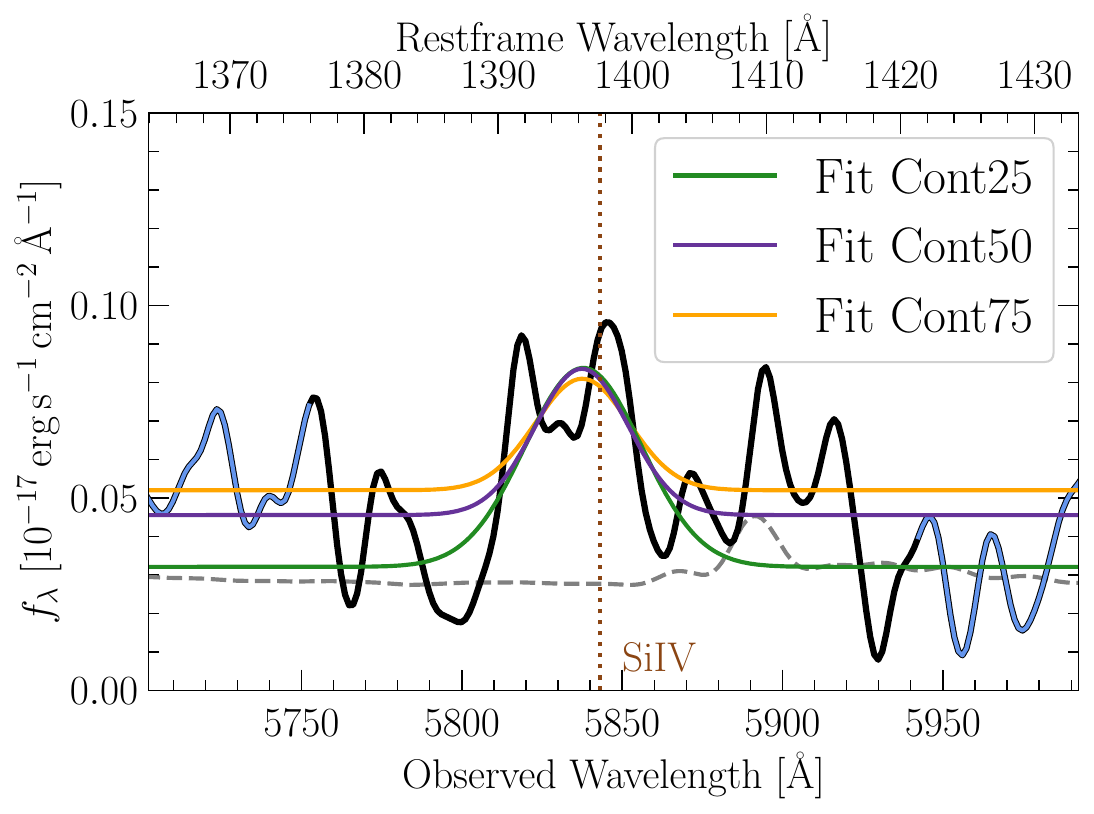}
   \includegraphics[width=0.49\hsize]{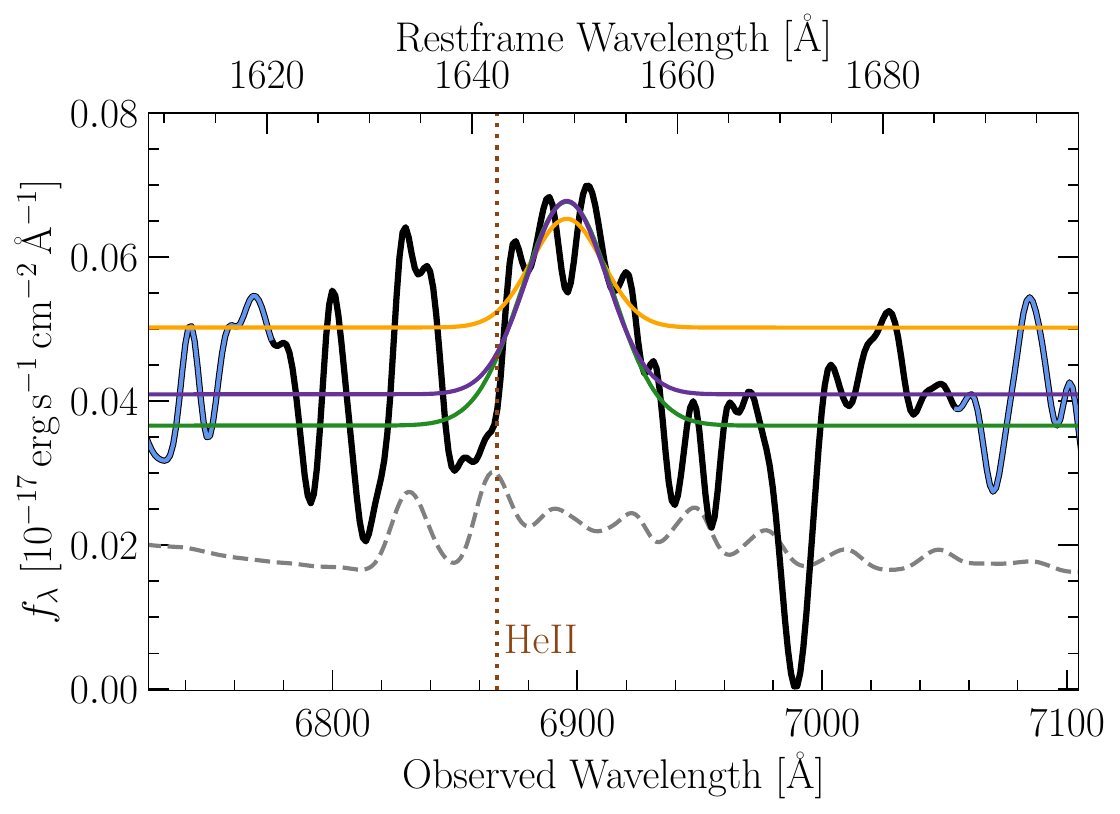}
   \includegraphics[width=0.49\hsize]{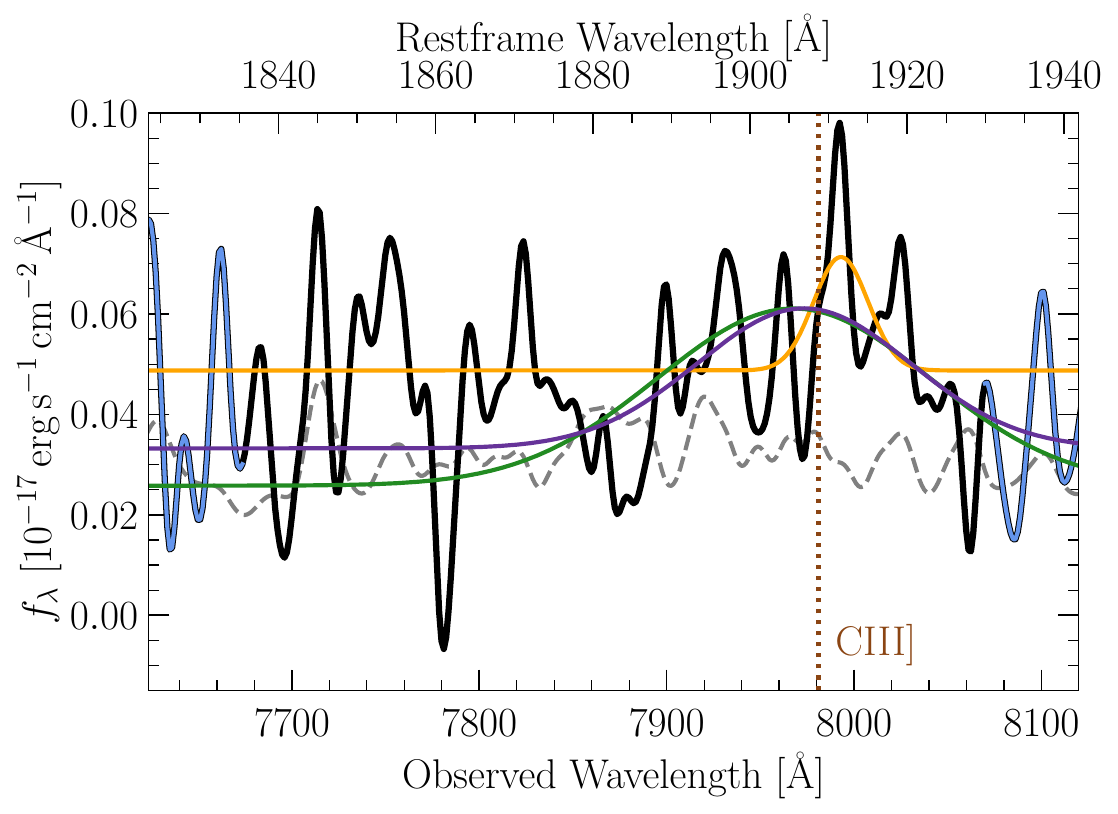}
      \caption{Gaussian fit to the emission lines after subtraction of the three different continua (Cont25, Cont50, Cont75) from the spectrum of AGNc1, smoothed by a 1D Gaussian kernel with $\sigma = 3$~pixels. The vertical dotted brown line indicates the expected wavelength $\lambda_{\rm sys}$ of the respective emission line, highlighting that the redshift determined from the peak of the \ciii\ line complex could be overestimated and the \heii\ line might be misidentified. Blue spectral regions mark the $50 \AA$ windows used to calculate the continuum.
      }
         \label{fig:fitagn2}
   \end{figure}

\begin{figure}
   \centering
   \includegraphics[width=0.49\hsize]{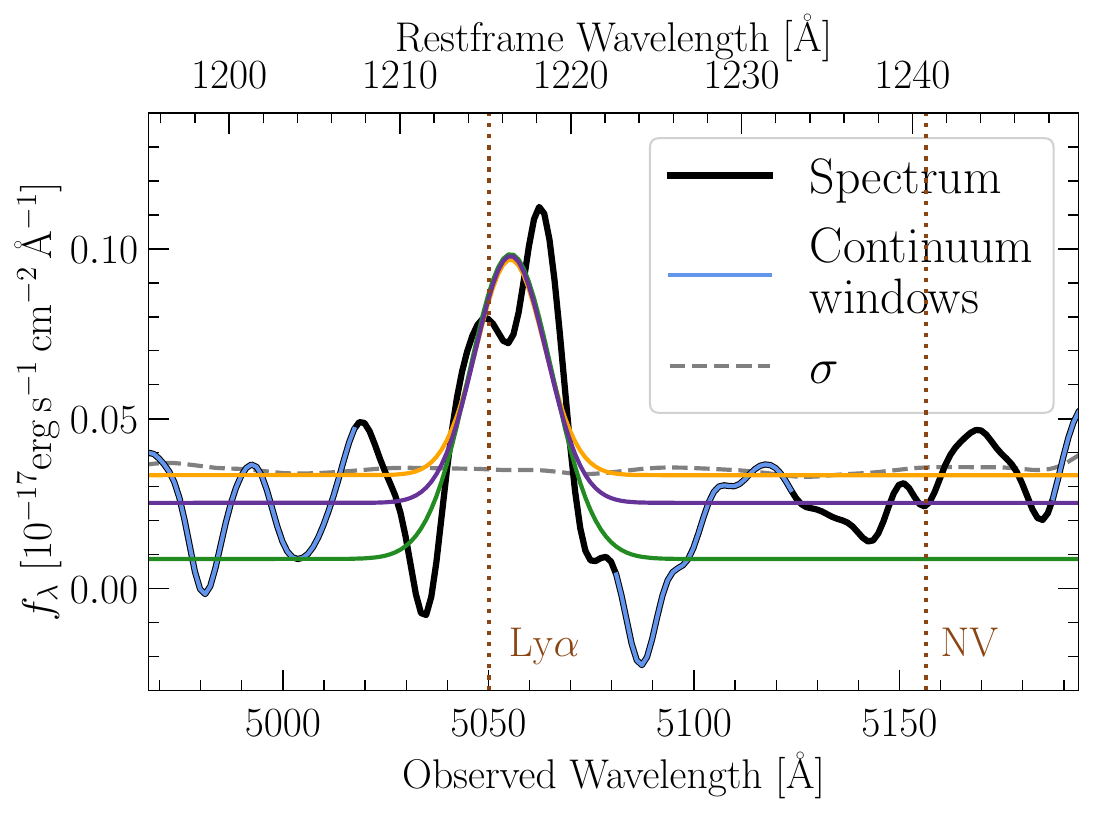}
   \includegraphics[width=0.49\hsize]{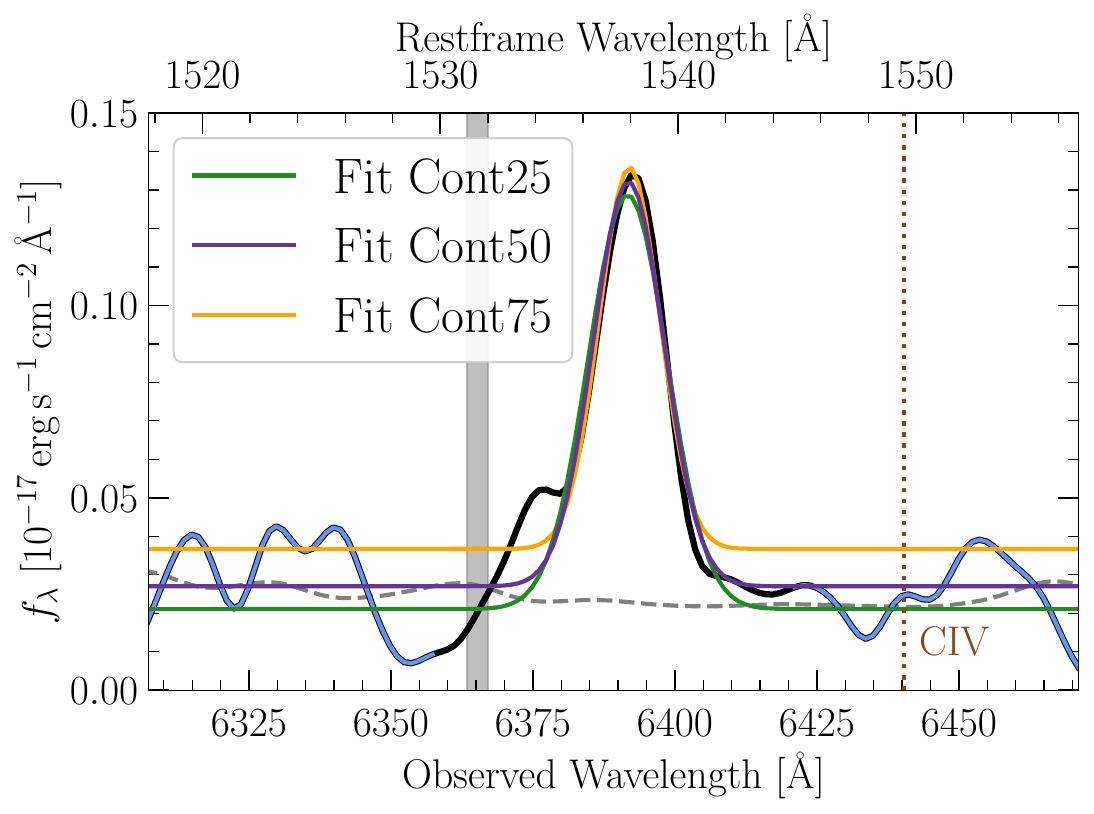}
   \includegraphics[width=0.49\hsize]{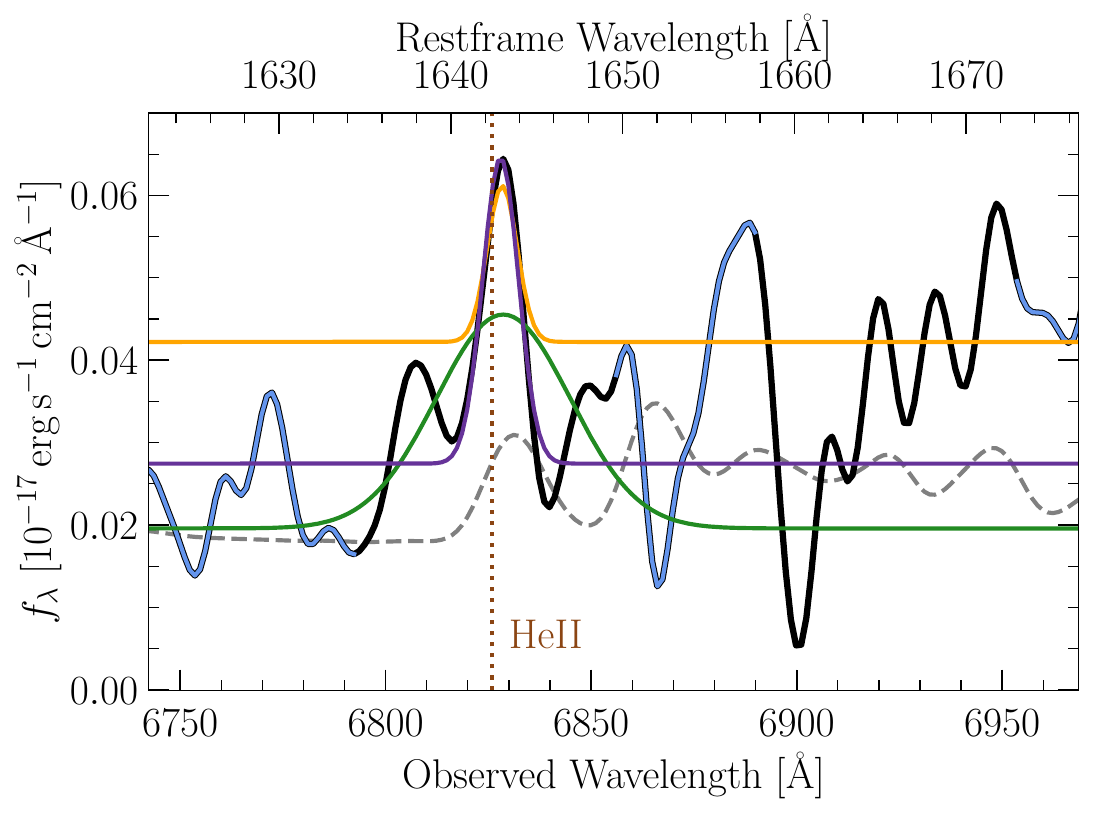}
   \includegraphics[width=0.49\hsize]{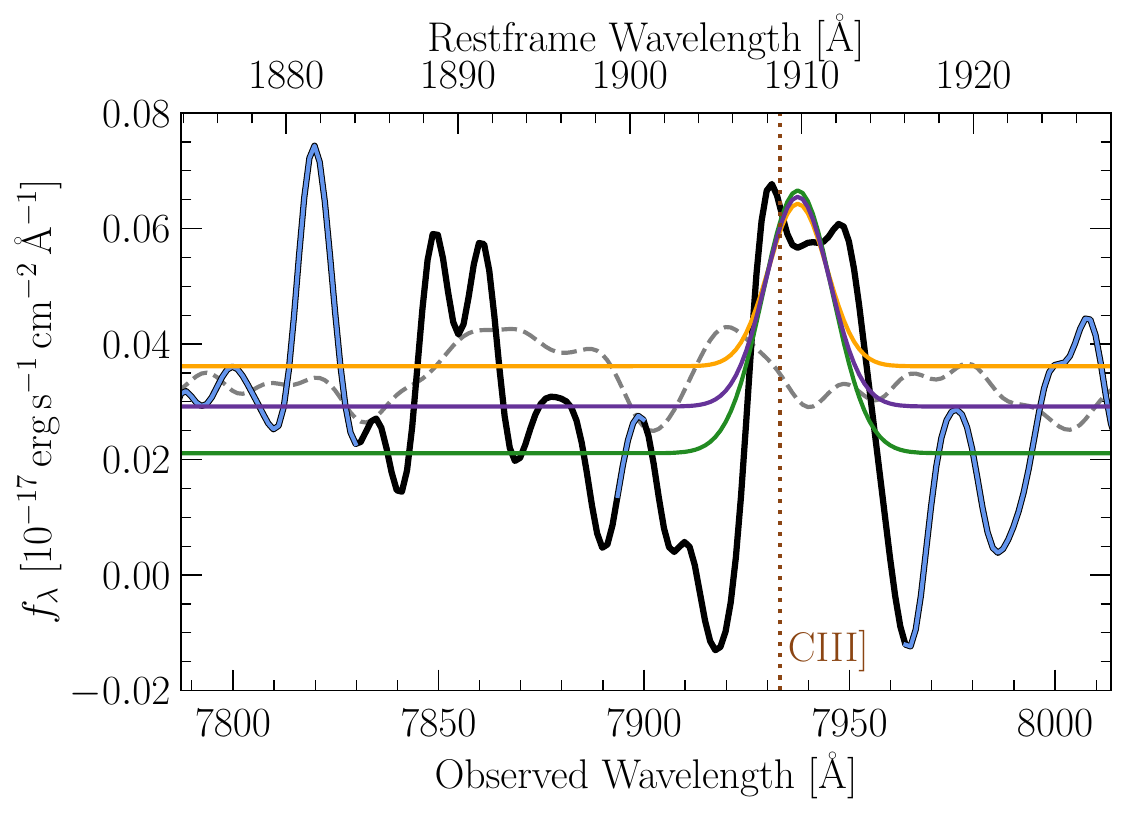}
      \caption{Same as Fig.~\ref{fig:fitagn2}, but for AGNc2. The \civ\ line peak of the fit is blueshifted with respect to $z_{\rm CIII]}$ by $-2208\ {\rm km\, s^{-1}}$. The shaded gray strip indicates the masked area affected by sky emission.
      }
         \label{fig:fitagn3}
   \end{figure}

Then, we calculated the continuum level and its variance by masking all shaded gray areas in Fig.~\ref{fig:spectra} as well as marked emission lines within $\pm 3 \sigma_{\rm ini}$ (AGNc1) and $\pm 5 \sigma_{\rm ini}$ (AGNc2) of the systemic wavelength. The chosen masking windows are large enough to ensure no line emission is contaminating the continuum calculation and small enough to retain local continuum around each fitted line. The latter is particularly important for the broad line width of AGNc1. Due to the large velocity shift of \civ\ in AGNc2, we adjusted the wavelength of the line downward by $10 \AA$ to ensure appropriate masking and fitting. For each fitted emission line, we then considered the next 40 unmasked wavelength channels (corresponding to 50~\AA) left and right of the expected peak $\lambda_{\rm sys}$ and calculated the 25th, 50th, and 75th percentile to get the median continuum value (Cont50) and typical range (Cont25, Cont75). Cont25 is consistent with the noise level for most lines (Fig.~\ref{fig:fitagn2},~\ref{fig:fitagn3}) and thus represents a conservative lower limit for the continuum level as the AGN candidates are detected in continuum emission (Fig.~\ref{fig:continuum}).
As the determination of the continuum introduces the largest uncertainty in the fit, in particular for the broad lines of AGNc1, we performed three emission line fittings using each of these continuum values for the subtraction and derived conservative line flux errors from the difference in results.

During line fitting, we assumed a single Gaussian model within the bounds given in Table~\ref{tab:bounds}, and weighted the fit with noise spectra extracted from the rescaled variance cube. The Ly$\alpha$ and \nv\ line of AGNc1 are likely blended together and are thus fitted together using two Gaussian lines centered on the respective systemic wavelengths.
From the Gaussian model of the \ciii\ line, we obtained the line width ${\rm FWHM_{CIII]}}$ of $5040^{+1390}_{-3790}$~km~s$^{-1}$ (AGNc1) and $600^{+80}_{-70}$~km~s$^{-1}$ (AGNc2). These values are larger than the calculation of the second moment in unsmoothed spectra, but are consistent with line widths in type~1 AGNs and type~2 AGNs respectively. We again corrected the line width of the Gaussian model by the smoothing kernel.

\begin{table}
\caption{Bounds of the Gaussian fitting model.}             
\label{tab:bounds}      
\centering          
\renewcommand{\arraystretch}{1.3}
\begin{tabular}{c c c c } 
\hline\hline       
Object & Normalization & Peak & $\sigma_{\rm Fit}$ \\ 
\hline                    
   AGNc1 & (0, inf) & $\lambda_{\rm sys}\, \pm \sigma_{\rm ini}$ & ($0.5\sigma_{\rm ini}$, $2.5\sigma_{\rm ini}$) \\
   AGNc2 & (0, inf) & $\lambda_{\rm sys}\, \pm \sigma_{\rm ini}$ & ($0.5\sigma_{\rm ini}$, $3\sigma_{\rm ini}$)  \\

\hline                  
\end{tabular}
\end{table}

If the flux value of an emission line is consistent with zero after subtracting Cont75, we consider it as undetected and only place upper limits on the flux. This is the case for \civ\ (AGNc1) and \nv\ and \siiv\ (AGNc2). We derived upper limits from the channel-wise RMS of the spectra scaled to $2.355\sigma_{\rm ini}$ and report the limit at $3\sigma$ significance.

We derived EWs and uncertainties for the high ionization lines \siiv\ and \civ\ from the three different Gaussian line fittings and their respective continuum levels and derived upper limits on EWs by assuming the median continuum level Cont50 and the upper flux limit.

\section{Estimating the efficiency of ram pressure stripping}
\label{sec:app1}
\begin{figure}[h]
   \centering
   \includegraphics[width=\hsize]{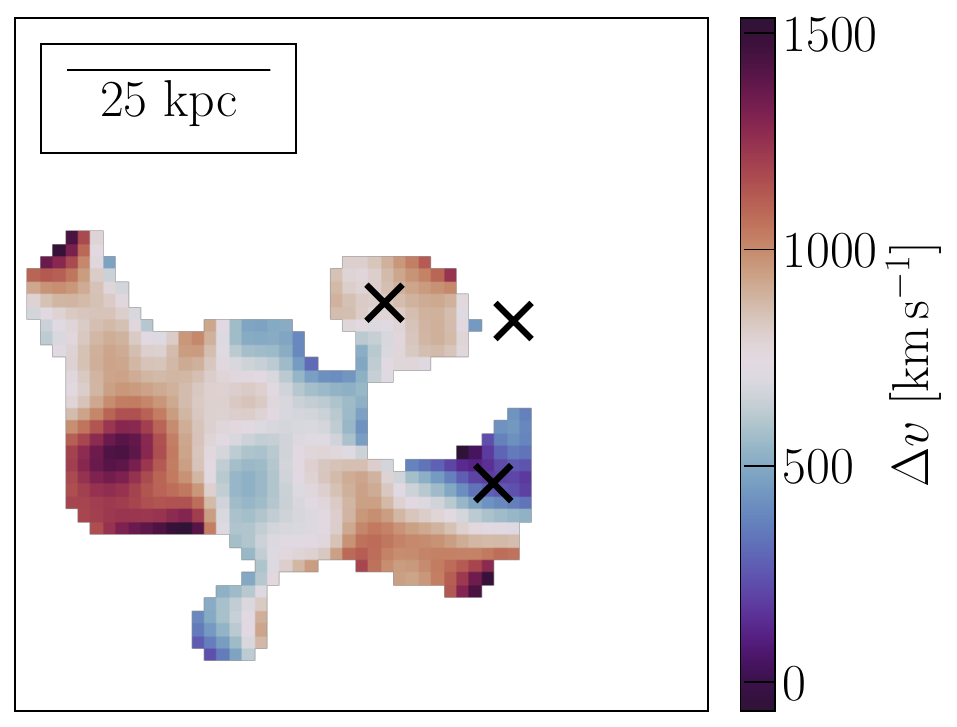}
      \caption{Moment 1 of the extended Ly$\alpha$ emission with respect to the systemic redshift of AGNc2. The centroid positions of QSO1, AGNc1 and AGNc2 are indicated with black crosses.
      }
         \label{fig:mom1}
   \end{figure}

The peculiar extended Ly$\alpha$ emission associated with the close triple system might be a further indication of on-going galaxy interaction as the emission could be a result of ram-pressure stripped ISM of AGNc2 appearing Ly$\alpha$ bright. This could further explain the large velocity differences in the nebula obtained from the first moment with respect to AGNc2 (Fig.~\ref{fig:mom1}), reaching 1500~km~s$^{-1}$ at the East edge of the isophote furthest away from the AGN candidate and generally approaching the systemic redshift of AGNc2 closer to it, although the nebula shows significant turbulence.
To determine the feasibility of this hypothesis, we estimated the efficiency of stripping in the configuration of the system.

Ram-pressure stripping is possible if $\rho_{\rm halo} v_{\rm AGNc2}^2 > 2 \pi G \sigma_s \sigma_g$ \citep{Gunn1972}.
In this, $\rho_{\rm halo}$ is the density of the hot gas in the halo of QSO1 and obtained from the typical volume-weighted number density in a hot halo of mass log($M$/M$_\odot) = 12.5 -13$ at 20~\% of the virial radius corresponding to the projected distance of AGNc2, ${\rm log}(n/[{\rm cm^{-3}}]) = -3.9$ \citep{Wijers2022}. This mass bin is chosen as it includes the halo mass estimate for single quasars at this redshift (${\rm 10^{12.5}\ M_{\odot}}$) as well as threefold this mass (${\rm 10^{13}\ M_{\odot}}$) in case QSO1, AGNc1 and AGNc2 all share one dark matter halo and all three resided in quasar-typical massive halos before. This larger halo mass is also similar to that estimated for one of the known triplets (\citealt{FAB:2022}). $v_{\rm AGNc2}$ is the velocity with which AGNc2 is crossing the halo of QSO1 and at least as high as the line-of-sight velocity reported in Table \ref{table:agn}. A conservative estimate of the 3D velocity is $v_{\rm AGNc2} = \sqrt{3} \Delta v_{\rm AGNc2} \approx 1318$~km~s$^{-1}$.

The gravitational pull of AGNc2 withstanding the ram pressure is described by the stellar and gas surface densities $\sigma_{\rm s}$ and $\sigma_{\rm g}$. AGN hosts at $z \sim 3$ typically have a massive ($M_{\star} \sim 10^{11} {\rm M_{\odot}}$), compact stellar component described by a Sersic profile with an average Sersic index of 2.6 and an effective radius of 0.9~kpc \citep{Kocevski2023}. The gas disk is well-described by an exponential profile, and we assumed an effective radius of 2~kpc \citep{Ward2024} and a cold gas mass fraction of 20~\%. 
The stripped gas mass can then be estimated by integrating the gas mass profile beyond the radius at which ram pressure exceeds the gravitational pull of the galaxy, and it amounts to at least $6.3 \times 10^9 {\rm M_{\odot}}$ when assuming the line-of-sight velocity, or roughly $8.6 \times 10^9 {\rm M_{\odot}}$ when using the 3D velocity.

To contrast this with the Ly$\alpha$-bright gas, we estimated the mass from the extended emission using the equation $M_{\rm cool} \sim A N_H m_p/X$ with the Ly$\alpha$ nebula area $A=1437$~kpc$^2$ \citep{Herwig2024}, the proton mass $m_p$ and hydrogen fraction $X = 0.76$ and the typical column density of cool gas in the circumgalactic medium, log($N_H/[{\rm cm^{-2}}]) = 20.5$ \citep{Lau2016}. The latter quantity is associated with the highest uncertainty as it can vary by one order of magnitude and it might not be applicable to extraordinary systems. 
With these values, we obtained an estimated cool gas mass in the Ly$\alpha$ nebula of $4.8 \times 10^9 {\rm M_{\odot}}$.

\section{Probability of finding a similarly close AGN triplet}
\label{sec:app2pcf}

The probability to find two additional AGNs close to a known quasar can be estimated using

\begin{displaymath}
    P \sim 4\ n_{\rm QSO}^2\ V\ \int_0^{r_{\rm max}}\ \xi (r) 4 \pi r^2 dr
\end{displaymath}

with the comoving number density of AGNs $n_{\rm QSO}$, the AGN-filled volume $V$, the maximum distance between the AGNs $r_{\rm max}$ and the two-point correlation function of quasars $\xi (r)$.
This follows from a similar argument as presented in \cite{Hennawi15}, but we do not include the forth object, QSO2, in the estimation as the large projected distance implies that QSO2 does not share a host halo with the other three objects (yet) and thus, other assumptions have to be made in the calculation. Specifically, in the case of a triplet, equation 2 in \citet{Hennawi15} will only contain three permutations for the two-point correlation function and one permutation for the three-point correlation function.
Although the three-point correlation function relates to $\xi$ as $\zeta \sim \xi^{\frac{3}{2}}$ (\citealt{Hennawi15}), in our order-of-magnitude estimation, we can approximate $\zeta \sim \xi$, hence the factor of four in the above probability equation. 

We inferred $n_{\rm QSO}$ from the quasar luminosity function at $z=3$ \citep{Shen20} with a lower magnitude limit of $m_{\rm min, 1450 \AA} = 25$~mag and the quasar bolometric luminosity correction from \cite{Runnoe2012} resulting in $n_{\rm QSO} = 9.3 \times 10^{-5}$cMpc$^{-3}$.
Due to the imprecision in redshift determinations from broad QSO emission lines and the possibility of high peculiar velocities causing additional Doppler shifts, we extrapolated the maximum distance in the triplet, $r_{\rm max}$, as the highest projected distance plus a 50~\% margin to account for projection effects. This yields a quasar-filled comoving volume of $V = 8 \times 10^{-3}$ cMpc$^3$.
On small scales, the two-point correlation function is well described with $\xi = \left( \frac{r}{r_0} \right)^{- \gamma}$ with $\gamma = 2$ and $r_0 = 5.4 \pm 0.3 h^{-1}$~cMpc \citep{Kayo2012}.
Plugging in these values, we obtained $P \approx 3 \times 10^{-8}$.

The candidate triple AGN presented in this work is closer than previously found AGN triples. To get an estimate of the probability to find any AGN triple on halo scales, we repeated the calculation using $r_{\rm max} = 150$~kpc, corresponding roughly to the virial radius of a halo with mass ${\rm 10^{13}\ M_{\odot}}$, and obtained a probability of $P \approx 2 \times 10^{-5}$.
In contrast, the probability $P_r$ to find three AGNs with $r_{\rm max} = 30$~kpc at random without considering clustering can be described with $P_r = n_{\rm QSO}^2 \times V^2$ \citep{Hennawi15} and would be exceedingly rare with $P_r = 5.5 \times 10^{-13}$.

\section{Examining possible sources of spectral contamination}
\label{sec:appcontamination}

Multiple sources of artifacts and spurious signal can contaminate observational data. Here, we examine a few of these effects and determine the level of contamination introduced.

\begin{figure}[h]
   \centering
   \includegraphics[width=\hsize]{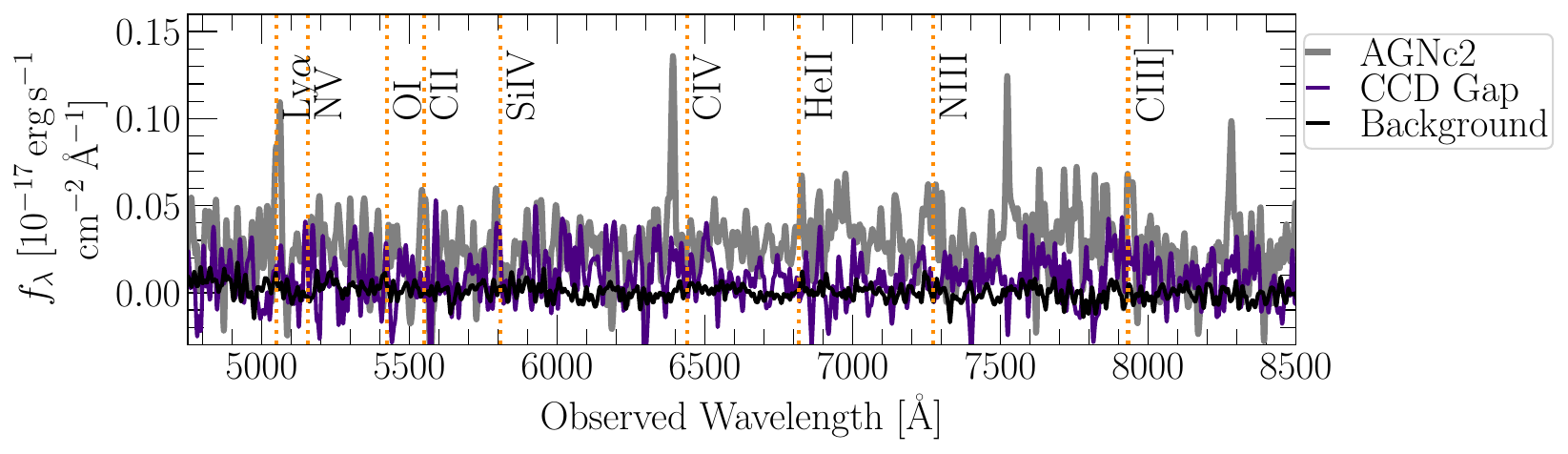}
      \caption{Check for possible sources of contamination in the spectrum of AGNc2. The spectrum taken from the region affected by CCD gap artifacts and the mean background spectrum are compared to the spectrum of AGNc2 and its expected emission line wavelengths.
      }
         \label{fig:ccdgap}
   \end{figure}

The first artifact to consider is residual emission from the CCD gap visible in the continuum image after subtraction of QSO1, AGNc1 and AGNc2 (Fig.~\ref{fig:continuum}, right panel). In particular, the spectrum of AGNc2 overlaps with the CCD gap. Thus, we extracted a spectrum at the same RA as AGNc2, but close to a flux peak on the CCD gap, and rescaled it to match the mean pixel value at the position of AGNc2 in the residual image. This region is indicated by a white plus in Fig.~\ref{fig:continuum}. The spectrum, smoothed with the same Gaussian kernel of size $\sigma = 3$~pixels, is displayed in Fig.~\ref{fig:ccdgap}. For reference, we also show the spectrum of AGNc2 as well as the expected wavelengths of emission lines.

While the CCD gap might introduce a noisy contribution to the continuum, it does not show strong emission features coinciding with the fitted emission lines of AGNc2.

Next, we examined the level of background emission in the cube by visually selecting ten regions within the same aperture as the spectra with a radius of 0.8$\arcsec$ without continuum emission, taking the mean of the extracted spectra and smoothing it with the Gaussian kernel of size $\sigma = 3$~pixels. The obtained spectrum is also displayed in Fig.~\ref{fig:ccdgap} and represents the typical channel-wise background level in the data. Due to its low level, it is not expected to contribute significantly to the continuum emission in any part of the spectra.

\begin{figure}[h]
   \centering
   \includegraphics[width=0.31\hsize]{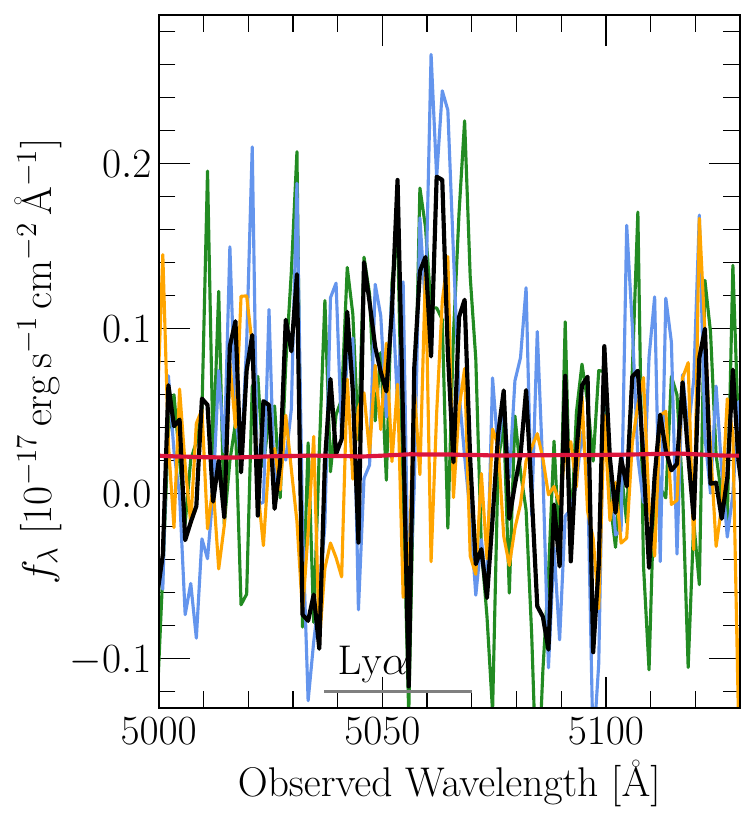}
   \includegraphics[width=0.33\hsize]{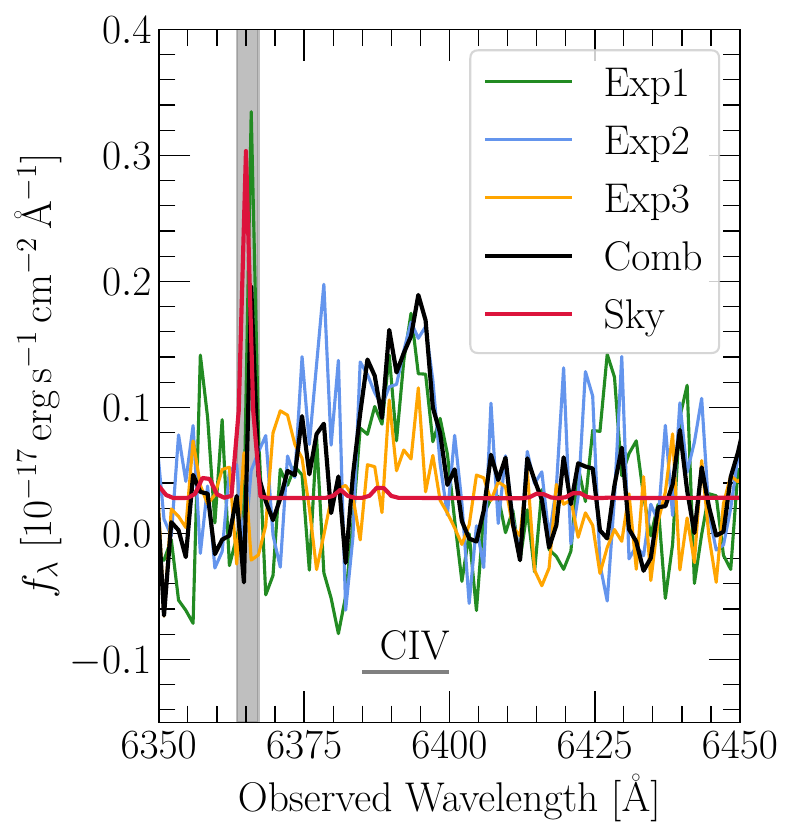}\\
   \includegraphics[width=0.33\hsize]{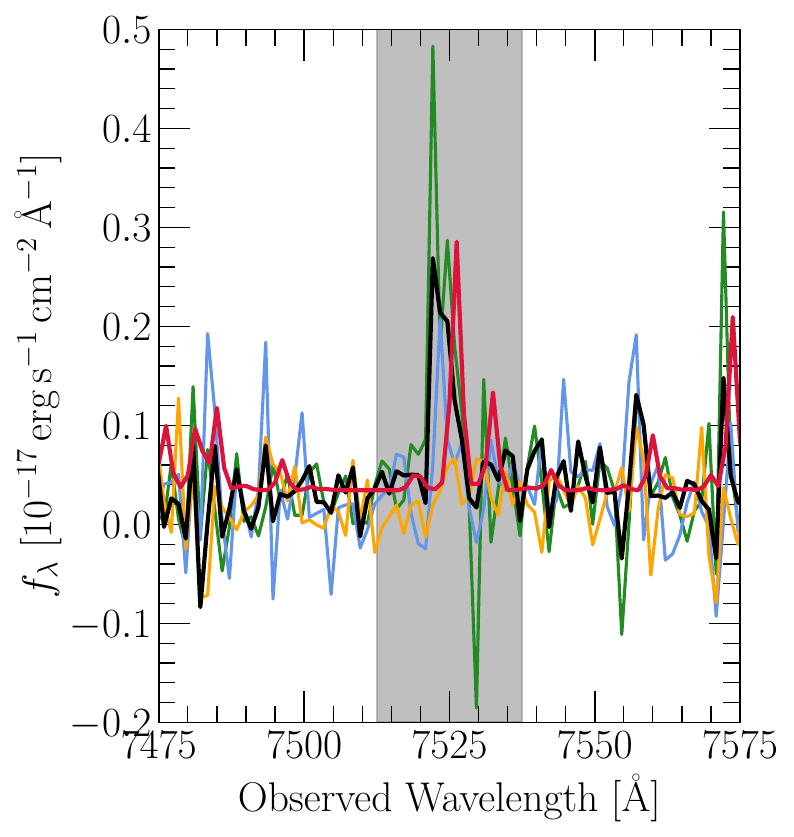}
   \includegraphics[width=0.31\hsize]{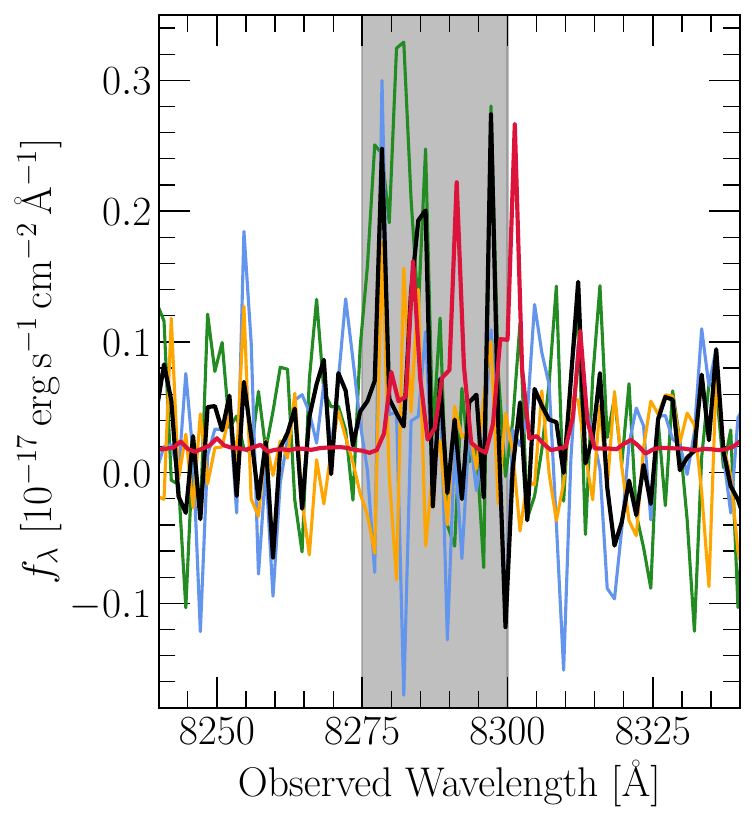}
      \caption{Unsmoothed spectra of AGNc2 extracted from the three exposures (Exp1, Exp2, Exp3, green, blue, yellow, respectively) and the combined cube (Comb, black) in spectral windows of narrow emission lines. A sky emission spectrum (Sky, red) is plotted in arbitrary flux units showing the position of atmospheric emission lines. Shaded gray areas are the same as in Fig.~\ref{fig:spectra}. To avoid contamination of the close-by \civ\ line, the sky emission line in the top right panel is masked before smoothing. The spectral extraction region falls onto the masked CCD gap in the third exposure, leading to a lower flux. We note that we performed the median combination of the cubes pixel by pixel, but summed the displayed spectra over multiple pixels. Consequently, the spectrum extracted from the median-combined cube differs from the median spectrum of those from the individual exposures.
      }
         \label{fig:skyemission}
   \end{figure}

A third source of contamination are strong atmospheric emission lines contributing significant flux peaks for the faint sources considered in this work. Despite careful subtraction of the sky spectrum in two different steps during the data reduction and before combining the individual exposures, residuals of lines can still remain in the combined data (\citealt{SOTO16}).
These residuals appear as very thin (one to three pixel wide) emission lines in the spectrum of continuum sources if they were masked together with source emission lines during sky removal. The spectrum of AGNc2 is specifically affected by this as there are multiple narrow lines visible in the unsmoothed spectrum of the combined cube as well as the individual exposures. In Fig.~\ref{fig:skyemission}, we show the affected spectral windows together with a sky emission model\footnote{Generated with the "The Cerro Paranal
Advanced Sky Model", \url{https://www.eso.org/observing/etc/doc/skycalc/The_Cerro_Paranal_Advanced_Sky_Model.pdf}}.

In these unsmoothed spectra, the difference in line width to \civ\ becomes apparent. Each of the narrow emission features not assignable to a common AGN UV line coincides with the position of a sky emission line within one or two pixels. Therefore, these additional lines are most likely sky subtraction residuals.

\begin{figure}[h]
   \centering
   \includegraphics[width=0.45\hsize]{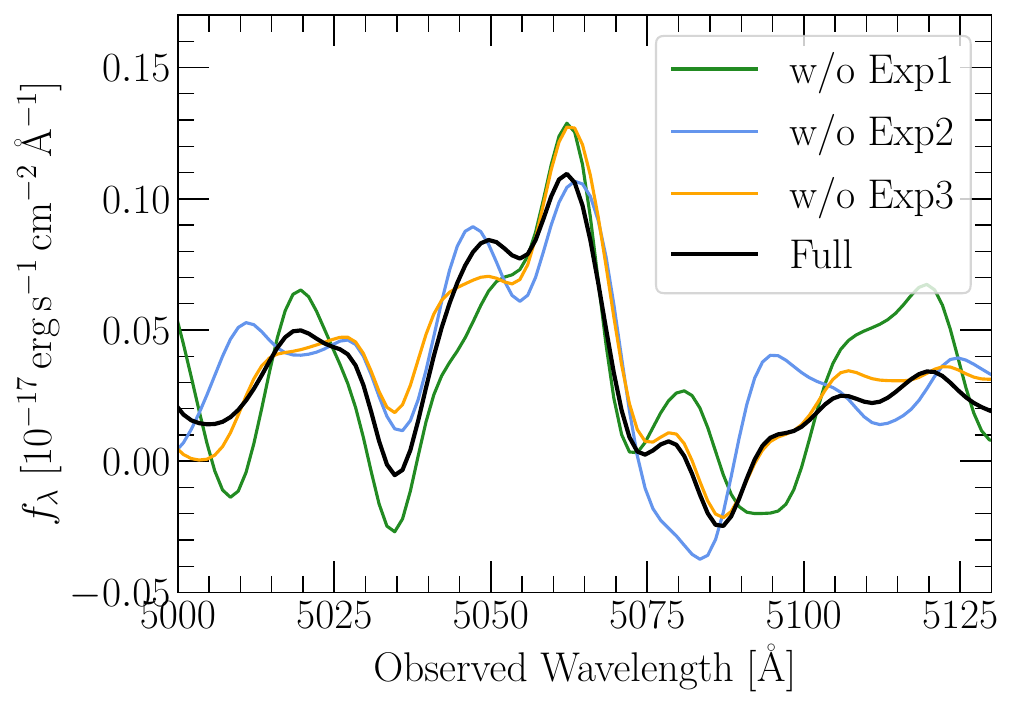}
   \includegraphics[width=0.45\hsize]{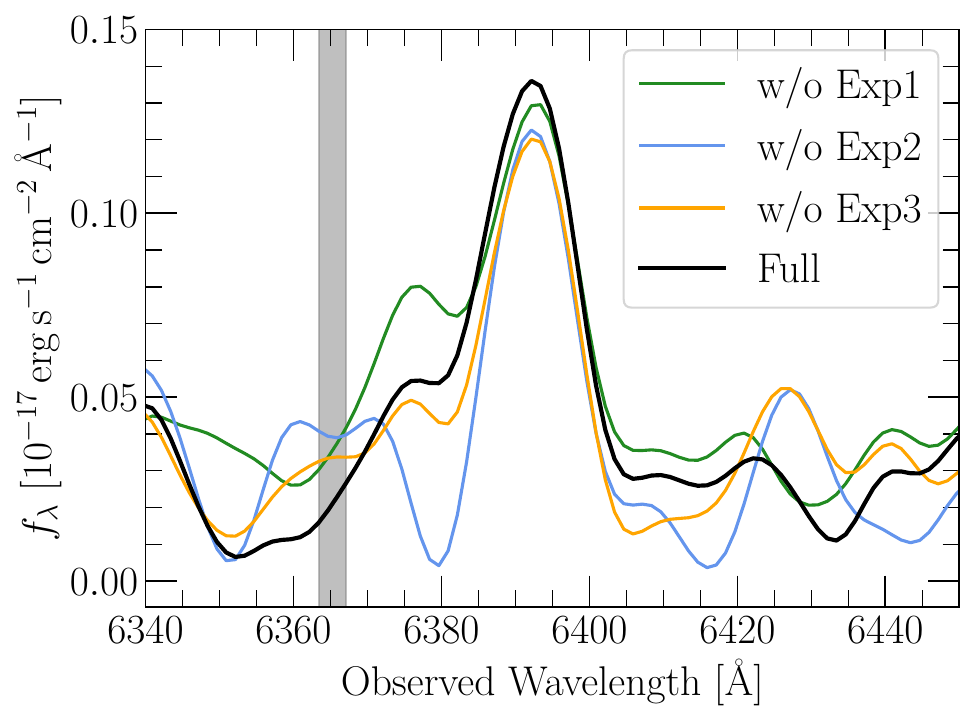}
      \caption{Smoothed spectra of AGNc2 at the wavelength of the Ly$\alpha$ (left) and \civ\ (right) emission extracted from cubes after combining only two of the three exposures as well as the full cube. The shaded gray area is the same as in Fig.~\ref{fig:spectra} and masked before smoothing.
      }
         \label{fig:combine}
   \end{figure}

Finally, given the small number of exposures obtained, individual datacubes could contribute significantly to emission lines, particularly if the line width is narrow. To exclude the possibility that the narrow lines in the spectrum of AGNc2 are spurious, we excluded in turn one of the three exposures during combination of the final cube and examined the Ly$\alpha$ and \civ\ lines in the different resulting spectra in comparison to the spectrum presented in Fig.~\ref{fig:spectra}. As shown in Fig.~\ref{fig:combine}, the line emission persists irrespective of the combination of exposures. We caution that the gap between CCDs, which is masked before combination of the cubes, covers parts of the PSF of AGNc2 in the third exposure. However, when combining only two cubes, masked areas in one data cube are filled in with the values of the other data cube. This is the reason why the seemingly decreased line flux of exposure 3 (Fig.~\ref{fig:skyemission}) does not have a strong affect on the line flux in the combined cube (Fig.~\ref{fig:combine}).

\end{appendix}
\end{document}